\documentclass[a4paper,11pt]{article}
\pdfoutput=1 
\usepackage[T1]{fontenc} 

\makeatletter
\gdef\@fpheader{ }
\gdef\@journal{ }
\makeatother
\RequirePackage{amsmath}
\RequirePackage{amssymb}
\RequirePackage{epsfig}
\RequirePackage{graphicx}
\RequirePackage[numbers,sort&compress]{natbib}
\RequirePackage{color}
\RequirePackage[colorlinks=true
,urlcolor=blue
,anchorcolor=blue
,citecolor=blue
,filecolor=blue
,linkcolor=blue
,menucolor=blue
,pagecolor=blue
,linktocpage=true
,pdfproducer=medialab
,pdfa=true
]{hyperref}

\newif\ifnotoc\notocfalse
\newif\ifemailadd\emailaddfalse
\newif\iftoccontinuous\toccontinuousfalse
\makeatletter
\def\@subheader{\@empty}
\def\@keywords{\@empty}
\def\@abstract{\@empty}
\def\@xtum{\@empty}
\def\@dedicated{\@empty}
\def\@arxivnumber{\@empty}
\def\@collaboration{\@empty}
\def\@collaborationImg{\@empty}
\def\@proceeding{\@empty}
\def\@preprint{\@empty}

\newcommand{\subheader}[1]{\gdef\@subheader{#1}}
\newcommand{\keywords}[1]{\if!\@keywords!\gdef\@keywords{#1}\else%
\PackageWarningNoLine{\jname}{Keywords already defined.\MessageBreak Ignoring last definition.}\fi}
\renewcommand{\abstract}[1]{\gdef\@abstract{#1}}
\newcommand{\dedicated}[1]{\gdef\@dedicated{#1}}
\newcommand{\arxivnumber}[1]{\gdef\@arxivnumber{#1}}
\newcommand{\proceeding}[1]{\gdef\@proceeding{#1}}
\newcommand{\xtumfont}[1]{\textsc{#1}}
\newcommand{\correctionref}[3]{\gdef\@xtum{\xtumfont{#1} \href{#2}{#3}}}
\newcommand\jname{JHEP}
\newcommand\acknowledgments{\section*{Acknowledgments}}

\newcommand\preprint[1]{\gdef\@preprint{\hfill #1}}

\makeatother

\newcommand\note[2][]{%
\if!#1!%
\stepcounter{footnote}\footnotetext{#2}%
\else%
{\renewcommand\thefootnote{#1}%
\footnotetext{#2}}%
\fi}

\makeatletter
\newtoks\auth@toks
\renewcommand{\author}[2][]{%
  \if!#1!%
    \auth@toks=\expandafter{\the\auth@toks#2\ }%
  \else
    \auth@toks=\expandafter{\the\auth@toks#2$^{#1}$\ }%
  \fi
}
\makeatother
\makeatletter
\newtoks\affil@toks\newif\ifaffil\affilfalse
\newcommand{\affiliation}[2][]{%
\affiltrue
  \if!#1!%
    \affil@toks=\expandafter{\the\affil@toks{\item[]#2}}%
  \else
    \affil@toks=\expandafter{\the\affil@toks{\item[$^{#1}$]#2}}%
  \fi
}
\makeatother
\makeatletter
\newtoks\email@toks\newcounter{email@counter}%
\newcommand{\emailAdd}[1]{%
\emailaddtrue%
\ifnum\theemail@counter>0\email@toks=\expandafter{\the\email@toks, \@email{#1}}%
\else\email@toks=\expandafter{\the\email@toks\@email{#1}}%
\fi\stepcounter{email@counter}}
\newcommand{\@email}[1]{\href{mailto:#1}{\tt #1}}
\makeatother

\makeatletter
\newcommand*\collaboration[1]{\gdef\@collaboration{#1}}
\newcommand*\collaborationImg[2][]{\gdef\@collaborationImg{#2}}
\makeatletter
\newcommand\afterLogoSpace{\smallskip}
\newcommand\afterSubheaderSpace{\vskip3pt plus 2pt minus 1pt}
\newcommand\afterProceedingsSpace{\vskip21pt plus0.4fil minus15pt}
\newcommand\afterTitleSpace{\vskip23pt plus0.06fil minus13pt}
\newcommand\afterRuleSpace{\vskip23pt plus0.06fil minus13pt}
\newcommand\afterCollaborationSpace{\vskip3pt plus 2pt minus 1pt}
\newcommand\afterCollaborationImgSpace{\vskip3pt plus 2pt minus 1pt}
\newcommand\afterAuthorSpace{\vskip5pt plus4pt minus4pt}
\newcommand\afterAffiliationSpace{\vskip3pt plus3pt}
\newcommand\afterEmailSpace{\vskip16pt plus9pt minus10pt\filbreak}
\newcommand\afterXtumSpace{\par\bigskip}
\newcommand\afterAbstractSpace{\vskip16pt plus9pt minus13pt}
\newcommand\afterKeywordsSpace{\vskip16pt plus9pt minus13pt}
\newcommand\afterArxivSpace{\vskip3pt plus0.01fil minus10pt}
\newcommand\afterDedicatedSpace{\vskip0pt plus0.01fil}
\newcommand\afterTocSpace{\bigskip\medskip}
\newcommand\afterTocRuleSpace{\bigskip\bigskip}
\newlength{\affiliationsSep}
\newcommand\beforetochook{\pagestyle{myplain}\pagenumbering{roman}}

\DeclareFixedFont\trfont{OT1}{phv}{b}{sc}{11}

\renewcommand\maketitle{
\pagestyle{empty}
\thispagestyle{titlepage}
\setcounter{page}{0}
\noindent{\small\scshape\@fpheader}\@preprint\par

\afterLogoSpace
\if!\@subheader!\else\noindent{\trfont{\@subheader}}\fi
\afterSubheaderSpace
\if!\@proceeding!\else\noindent{\sc\@proceeding}\fi
\afterProceedingsSpace
{\LARGE\flushleft\sffamily\bfseries\@title\par}
\afterTitleSpace
\hrule height 1.5\p@%
\afterRuleSpace
\if!\@collaboration!\else
{\Large\bfseries\sffamily\raggedright\@collaboration}\par
\afterCollaborationSpace
\fi
\if!\@collaborationImg!\else
{\normalsize\bfseries\sffamily\raggedright\@collaborationImg}\par
\afterCollaborationImgSpace
\fi
{\bfseries\raggedright\sffamily\the\auth@toks\par}
\afterAuthorSpace
\ifaffil\begin{list}{}{%
\setlength{\leftmargin}{0.28cm}%
\setlength{\labelsep}{0pt}%
\setlength{\itemsep}{\affiliationsSep}%
\setlength{\topsep}{-\parskip}}
\itshape\small%
\the\affil@toks
\end{list}\fi
\afterAffiliationSpace
\ifemailadd 
\noindent\hspace{0.28cm}\begin{minipage}[l]{.9\textwidth}
\begin{flushleft}
\textit{E-mail:} \the\email@toks
\end{flushleft}
\end{minipage}
\else 
\PackageWarningNoLine{\jname}{E-mails are missing.\MessageBreak Plese use \protect\emailAdd\space macro to provide e-mails.}
\fi
\afterEmailSpace
\if!\@xtum!\else\noindent{\@xtum}\afterXtumSpace\fi
\if!\@abstract!\else\noindent{\renewcommand\baselinestretch{.9}\textsc{Abstract:}}\ \@abstract\afterAbstractSpace\fi
\if!\@keywords!\else\noindent{\textsc{Keywords:}} \@keywords\afterKeywordsSpace\fi
\if!\@arxivnumber!\else\noindent{\textsc{ArXiv ePrint:}} \href{http://arxiv.org/abs/\@arxivnumber}{\@arxivnumber}\afterArxivSpace\fi
\if!\@dedicated!\else\vbox{\small\it\raggedleft\@dedicated}\afterDedicatedSpace\fi
\ifnotoc\else
\iftoccontinuous\else\newpage\fi
\beforetochook\hrule
\tableofcontents
\afterTocSpace
\hrule
\afterTocRuleSpace
\fi
\setcounter{footnote}{0}
\pagestyle{myplain}\pagenumbering{arabic}
} 

\renewcommand{\baselinestretch}{1.1}\normalsize
\divide\@tempdima\baselineskip
\@tempcnta=\@tempdima
\skip\@mpfootins = \skip\footins
\renewcommand{\@dotsep}{10000}

\newcommand\ps@myplain{
\pagenumbering{arabic}
\renewcommand\@oddfoot{\hfill-- \thepage\ --\hfill}
\renewcommand\@oddhead{}}
\let\ps@plain=\ps@myplain

\newcommand\ps@titlepage{\renewcommand\@oddfoot{}\renewcommand\@oddhead{}}

\numberwithin{equation}{section}

\renewcommand\section{\@startsection{section}{1}{\z@}%
                                   {-3.5ex \@plus -1.3ex \@minus -.7ex}%
                                   {2.3ex \@plus.4ex \@minus .4ex}%
                                   {\normalfont\large\bfseries}}
\renewcommand\subsection{\@startsection{subsection}{2}{\z@}%
                                   {-2.3ex\@plus -1ex \@minus -.5ex}%
                                   {1.2ex \@plus .3ex \@minus .3ex}%
                                   {\normalfont\normalsize\bfseries}}
\renewcommand\subsubsection{\@startsection{subsubsection}{3}{\z@}%
                                   {-2.3ex\@plus -1ex \@minus -.5ex}%
                                   {1ex \@plus .2ex \@minus .2ex}%
                                   {\normalfont\normalsize\bfseries}}
\renewcommand\paragraph{\@startsection{paragraph}{4}{\z@}%
                                   {1.75ex \@plus1ex \@minus.2ex}%
                                   {-1em}%
                                   {\normalfont\normalsize\bfseries}}
\renewcommand\subparagraph{\@startsection{subparagraph}{5}{\parindent}%
                                   {1.75ex \@plus1ex \@minus .2ex}%
                                   {-1em}%
                                   {\normalfont\normalsize\bfseries}}

\def\fnum@figure{\textbf{\figurename\nobreakspace\thefigure}}
\def\fnum@table{\textbf{\tablename\nobreakspace\thetable}}

\long\def\@makecaption#1#2{%
  \vskip\abovecaptionskip
  \sbox\@tempboxa{\small #1. #2}%
  \ifdim \wd\@tempboxa >\hsize
    \small #1. #2\par
  \else
    \global \@minipagefalse
    \hb@xt@\hsize{\hfil\box\@tempboxa\hfil}%
  \fi
  \vskip\belowcaptionskip}

\renewenvironment{thebibliography}[1]{%
\begin{oldthebibliography}{#1}%
\small%
\raggedright%
\setlength{\itemsep}{5pt plus 0.2ex minus 0.05ex}%
}%
{%
\end{oldthebibliography}%
}

\begin{document}


\title{\boldmath Scattering theory without large-distance asymptotics}

\author[a]{Tong Liu,}
\author[a]{Wen-Du Li,}
\author[a,b,1]{and Wu-Sheng Dai}\note{daiwusheng@tju.edu.cn.}


\affiliation[a]{Department of Physics, Tianjin University, Tianjin 300072, P.R. China}
\affiliation[b]{LiuHui Center for Applied Mathematics, Nankai University \& Tianjin University, Tianjin 300072, P.R. China}








\abstract{In conventional scattering theory, to obtain an explicit result, one imposes a
precondition that the distance between target and observer is infinite. With
the help of this precondition, one can asymptotically replace the Hankel
function and the Bessel function with the sine functions so that one can
achieve an explicit result. Nevertheless, after such a treatment, the
information of the distance between target and observer is inevitably lost. In
this paper, we show that such a precondition is not necessary: without losing
any information of distance, one can still obtain an explicit result of a
scattering rigorously. In other words, we give an rigorous explicit scattering
result which contains the information of distance between target and observer.
We show that at a finite distance, a modification factor --- the Bessel
polynomial --- appears in the scattering amplitude, and, consequently, the
cross section depends on the distance, the outgoing wave-front surface is no
longer a sphere, and, besides the phase shift, there is an additional phase
(the argument of the Bessel polynomial) appears in the scattering wave function.}

\maketitle
\flushbottom


\section{Introduction}

In conventional scattering theory, which is now a standard quantum mechanics
textbook content, to seek an explicit result, one imposes a precondition that
the distance between target and observer is infinite. As a result, the
conventional scattering theory loses all the information of the distance and
the result depends only on the angle of emergence. In this paper, we will show
that without such a precondition, one can still achieve a rigorous scattering
theory which, of course, contains the information of distance that is lost in
conventional scattering theory.

The dynamical information of a scattering problem with a spherical potential
$V\left(  r\right)  $ are embedded in the radial wave equation,%
\begin{equation}
\frac{1}{r^{2}}\frac{d}{dr}\left(  r^{2}\frac{dR_{l}}{dr}\right)  +\left[
k^{2}-\frac{l\left(  l+1\right)  }{r^{2}}-V\left(  r\right)  \right]  R_{l}=0.
\label{ratial equation}%
\end{equation}
The scattering boundary condition in conventional scattering theory is taken
to be
\begin{equation}
\psi\left(  r,\theta\right)  =e^{ikr\cos\theta}+f\left(  \theta\right)
\frac{e^{ikr}}{r},\text{ \ }r\rightarrow\infty. \label{bcinf}%
\end{equation}

In conventional scattering theory, in order to achieve an explicit result, two
kinds of asymptotic approximations are employed.

1) Replace the solution of the free radial equation, i.e., Eq.
(\ref{ratial equation}) with $V\left(  r\right)  =0$, with its asymptotics:%
\begin{align}
R_{l}\left(  r\right)   &  =C_{l}h_{l}^{\left(  2\right)  }\left(  kr\right)
+D_{l}h_{l}^{\left(  1\right)  }\left(  kr\right) \label{asyRl}\\
&  \overset{r\rightarrow\infty}{\sim}A_{l}\frac{\sin\left(  kr-l\pi
/2+\delta_{l}\right)  }{kr}, \label{asyRlinf}%
\end{align}
where $h_{l}^{\left(  1\right)  }\left(  z\right)  $ and $h_{l}^{\left(
2\right)  }\left(  z\right)  $ are the first and second kind spherical Hankel
functions, $e^{2i\delta_{l}}=D_{l}/C_{l}$ defines the scattering phase shift
$\delta_{l}$, and $A_{l}=2\sqrt{C_{l}D_{l}}$.

2) Replace the plane wave expansion in the boundary condition with its
asymptotics:%
\begin{align}
e^{ikr\cos\theta}  &  =\sum_{l=0}^{\infty}\left(  2l+1\right)  i^{l}%
j_{l}\left(  kr\right)  P_{l}\left(  \cos\theta\right) \label{pmbexp}\\
&  \overset{r\rightarrow\infty}{\sim}\sum_{l=0}^{\infty}\left(  2l+1\right)
i^{l}\frac{\sin\left(  kr-l\pi/2\right)  }{kr}P_{l}\left(  \cos\theta\right)
, \label{pmbasym}%
\end{align}
where $j_{l}\left(  z\right)  $ is the spherical Bessel function.

Technologically speaking, the above two treatments in conventional theory are
to replace the spherical Hankel function, $h_{l}^{\left(  1\right)  }\left(
kr\right)  $ and $h_{l}^{\left(  2\right)  }\left(  kr\right)  $, and the
spherical Bessel function, $j_{l}\left(  kr\right)  $, with their asymptotics,
and, thus, inevitably lead to the loss of information of the distance $r$.

In this paper, we will show that the above two replacements is not necessary;
without these two replacements, we can still obtain a rigorous scattering
theory which contains the information of the distance between target and observer.

A systematic rigorous result of a scattering with the distance between target
and observer is given in Sec. \ref{rigorousresults}. The conclusion and
outlook are given in Sec. \ref{conclusions}.

\section{ Rigorous result of scattering without large-distance asymptotics
\label{rigorousresults}}

In this section, a rigorous treatment without large-distance asymptotics for
short-range potentials is established. The scattering wave function,
scattering amplitude, phase shift, cross section, and a description of the
outgoing wave are rigorously obtained.

\subsection{Phase shift}

In conventional scattering theory, as mentioned above, one replaces the
solution of the free radial equation, $R_{l}\left(  r\right)  $, given by Eq.
(\ref{asyRl}) with its asymptotics, Eq. (\ref{asyRlinf}), using the
asymptotics of the spherical Hankel functions $h_{l}^{\left(  1\right)
}\left(  kr\right)  \sim$ $\frac{1}{ikr}e^{i\left(  kr-l\pi/2\right)  }$ and
$h_{l}^{\left(  2\right)  }\left(  kr\right)  \sim$ $-\frac{1}{ikr}%
e^{-i\left(  kr-l\pi/2\right)  }$. Obviously, such a replacement will lose information.

In the following, with $R_{l}\left(  r\right)  $ given by Eq. (\ref{asyRl}),
rather than its asymptotics, Eq. (\ref{asyRlinf}), we solve the scattering rigorously.

The first step is to rewrite $R_{l}\left(  r\right)  $ given by Eq.
(\ref{asyRl}) as
\begin{align}
R_{l}\left(  r\right)   &  =C_{l}h_{l}^{\left(  2\right)  }\left(  kr\right)
+D_{l}h_{l}^{\left(  1\right)  }\left(  kr\right) \nonumber\\
&  =M_{l}\left(  \frac{i}{kr}\right)  \frac{A_{l}}{kr}\sin\left[
kr-\frac{l\pi}{2}+\delta_{l}+\Delta_{l}\left(  \frac{i}{kr}\right)  \right]  ,
\label{Rlr}%
\end{align}
where $e^{2i\delta_{l}}=D_{l}/C_{l}$ and $M_{l}\left(  x\right)  =\left\vert
y_{l}\left(  x\right)  \right\vert $ and $\Delta_{l}\left(  x\right)  =\arg
y_{l}\left(  x\right)  $ are the modulus and argument of the Bessel polynomial
$y_{l}\left(  x\right)  $, respectively.

In order to achieve Eq. (\ref{Rlr}), we prove the relation%
\begin{equation}
C_{l}h_{l}^{\left(  2\right)  }\left(  x\right)  +D_{l}h_{l}^{\left(
1\right)  }\left(  x\right)  =M_{l}\left(  \frac{i}{x}\right)  \frac{A_{l}}%
{x}\sin\left[  x-\frac{l\pi}{2}+\delta_{l}+\Delta_{l}\left(  \frac{i}%
{x}\right)  \right]  .\label{flx}%
\end{equation}

\textit{Proof. }The\textit{ }first and second kind spherical Hankel functions,
$h_{l}^{\left(  1\right)  }\left(  x\right)  $ and $h_{l}^{\left(  2\right)
}\left(  x\right)  $, can be expanded as \cite{abramowitz1964handbook}%
\begin{align}
h_{l}^{\left(  1\right)  }\left(  x\right)   &  =e^{ix}\sum_{k=0}^{l}%
\frac{i^{k-l-1}\left(  l+k\right)  !}{2^{k}k!\left(  l-k\right)  !x^{k+1}},\\
h_{l}^{\left(  2\right)  }\left(  x\right)   &  =e^{-ix}\sum_{k=0}^{l}%
\frac{\left(  -i\right)  ^{k-l-1}\left(  l+k\right)  !}{2^{k}k!\left(
l-k\right)  !x^{k+1}}.
\end{align}
By the Bessel polynomial \cite{abramowitz1964handbook},
\begin{equation}
y_{l}\left(  x\right)  =\sum_{k=0}^{l}\frac{\left(  l+k\right)  !}{k!\left(
l-k\right)  !}\left(  \frac{x}{2}\right)  ^{k}, \label{BP}%
\end{equation}
we can rewrite $h_{l}^{\left(  1\right)  }\left(  x\right)  $ and
$h_{l}^{\left(  2\right)  }\left(  x\right)  $ as%
\begin{align}
h_{l}^{\left(  1\right)  }\left(  x\right)   &  =e^{i\left(  x-l\pi/2\right)
}\frac{1}{ix}y_{l}\left(  \frac{i}{x}\right)  ,\nonumber\\
h_{l}^{\left(  2\right)  }\left(  x\right)   &  =-e^{-i\left(  x-l\pi
/2\right)  }\frac{1}{ix}y_{l}\left(  -\frac{i}{x}\right)  . \label{hy}%
\end{align}
Using Eq. (\ref{hy}), we have%
\begin{equation}
C_{l}h_{l}^{\left(  2\right)  }\left(  x\right)  +D_{l}h_{l}^{\left(
1\right)  }\left(  x\right)  =C_{l}\left[  -\frac{e^{-i\left(  x-l\pi
/2\right)  }}{ix}y_{l}\left(  -\frac{i}{x}\right)  +e^{2i\delta_{l}}%
\frac{e^{i\left(  x-l\pi/2\right)  }}{ix}y_{l}\left(  \frac{i}{x}\right)
\right]  . \label{flx1}%
\end{equation}
Writing the Bessel polynomial as $y_{l}=M_{l}e^{i\Delta_{l}}$, we prove the
relation (\ref{flx}).

The wave function, then, by $\psi\left(  r,\theta\right)  =\sum_{l=0}^{\infty
}R_{l}\left(  r\right)  P_{l}\left(  \cos\theta\right)  $, can be obtained
immediately from Eq. (\ref{Rlr}),%
\begin{equation}
\psi\left(  r,\theta\right)  =\sum_{l=0}^{\infty}M_{l}\left(  \frac{i}%
{kr}\right)  \frac{A_{l}}{kr}\sin\left[  kr-\frac{l\pi}{2}+\delta_{l}%
+\Delta_{l}\left(  \frac{i}{kr}\right)  \right]  P_{l}\left(  \cos
\theta\right)  . \label{psiasy}%
\end{equation}

When the distance $r$ is finite,\ the coefficient becomes $M_{l}A_{l}$ and the
phase becomes $\delta_{l}+\Delta_{l}$, where $M_{l}$ and $\Delta_{l}$ both
depend on $r$. While, in conventional scattering theory, $r\rightarrow\infty$,
the coefficient is $A_{l}$ and the phase is $\delta_{l}$, and they are both
independent of $r$.

It should be emphasized that $\delta_{l}$ here is the same as that in
conventional scattering theory. This is because $\delta_{l}\ $is determined
only by the coefficient $C_{l}$ and $D_{l}$ and $y_{l}\left(  \frac{i}%
{kr}\right)  \overset{r\rightarrow\infty}{=}1$. Thus when $r\rightarrow\infty
$, $C_{l}$, $D_{l}$, and, accordingly, $\delta_{l}$ remains unchanged.

The modification factors, $\Delta_{l}$ and $M_{l}$, are independent of
potentials. When $r\rightarrow\infty$, $M_{l}\left(  r\rightarrow
\infty\right)  =1$ and $\Delta_{l}\left(  r\rightarrow\infty\right)  =0$.

\subsection{Asymptotic boundary condition}

The outgoing wave is no longer a spherical wave when the observer stands at a
finite distance from the target, other than that in large-distance
asymptotics. The outgoing wave now becomes a surface of revolution around the
incident direction, determined by the potential and the observation distance.
Because the outgoing waves are different at different distances, there is no
uniform expression of the asymptotic boundary condition like Eq.
(\ref{bcinf}). Here, we express the boundary condition as%

\begin{equation}
\psi\left(  r,\theta\right)  =e^{ikr\cos\theta}+f\left(  r,\theta\right)
\frac{e^{ikr}}{r}, \label{asymbc}%
\end{equation}
where $f\left(  r,\theta\right)  $ depends not only on $\theta$ but also on
$r$.

When the distance $r$ is finite, however, the differential scattering cross
section is no longer the square modulus of $f\left(  r,\theta\right)  $. Only
when $r\rightarrow\infty$, $f\left(  \infty,\theta\right)  =f\left(
\theta\right)  $ and the differential cross section reduces to $\left\vert
f\left(  \theta\right)  \right\vert ^{2}$.

To calculate $f\left(  r,\theta\right)  $, as that in conventional scattering
theory, we expand the incoming plane wave $e^{ikr\cos\theta}$ by the
eigenfunction of the angular momentum. Now, we prove that the expansion of
$e^{ikr\cos\theta}$, Eq. (\ref{pmbexp}), can be\ exactly rewritten as%
\begin{align}
e^{ikr\cos\theta}  &  =\sum_{l=0}^{\infty}\left(  2l+1\right)  i^{l}%
j_{l}\left(  kr\right)  P_{l}\left(  \cos\theta\right) \nonumber\\
&  =\sum_{l=0}^{\infty}\left(  2l+1\right)  i^{l}M_{l}\left(  \frac{i}%
{kr}\right)  \frac{1}{kr}\sin\left[  kr-\frac{l\pi}{2}+\Delta_{l}\left(
\frac{i}{kr}\right)  \right]  P_{l}\left(  \cos\theta\right)  .
\label{ntpwexp}%
\end{align}

\textit{Proof. }A plane wave can be expanded as \cite{joachain1975quantum}
\begin{equation}
e^{ikr\cos\theta}=\sum_{l=0}^{\infty}\left(  2l+1\right)  i^{l}j_{l}\left(
kr\right)  P_{l}\left(  \cos\theta\right)  . \label{pwexp}%
\end{equation}
By the relations $h_{l}^{\left(  1\right)  }\left(  x\right)  =j_{l}\left(
x\right)  +in_{l}\left(  x\right)  $ and $h_{l}^{\left(  2\right)  }\left(
x\right)  =j_{l}\left(  x\right)  -in_{l}\left(  x\right)  $, the spherical
Bessel function $j_{l}\left(  x\right)  $ can be rewritten as $j_{l}\left(
x\right)  =\frac{1}{2}\left[  h_{l}^{\left(  1\right)  }\left(  x\right)
+h_{l}^{\left(  2\right)  }\left(  x\right)  \right]  $, where $n_{l}\left(
x\right)  $ is the spherical Neumann function \cite{abramowitz1964handbook}.
By Eq. (\ref{hy}), we have
\begin{equation}
j_{l}\left(  kr\right)  =M_{l}\left(  \frac{i}{kr}\right)  \frac{1}{kr}%
\sin\left[  kr-\frac{l\pi}{2}+\Delta_{l}\left(  \frac{i}{kr}\right)  \right]
. \label{jl-sin}%
\end{equation}
Substituting this result into Eq. (\ref{pwexp}) proves Eq. (\ref{ntpwexp}).

The plane wave expansion (\ref{ntpwexp}) is exact, rather than the asymptotic
one, Eq. (\ref{pmbasym}), used in conventional scattering theory. In
conventional scattering theory, the spherical Bessel function $j_{l}\left(
kr\right)  $ given by Eq. (\ref{jl-sin}) is replaced by its asymptotics:
$j_{l}\left(  kr\right)  \sim\frac{1}{kr}\sin\left(  kr-l\pi/2\right)  $,
i.e., $M_{l}$ and $\Delta_{l}$ are asymptotically taken to be $M_{l}\left(
\frac{i}{kr}\right)  \sim1$ and $\Delta_{l}\left(  \frac{i}{kr}\right)  \sim
0$; as a result, the information embedded in $M_{l}$ and $\Delta_{l}$ is lost.

The boundary condition, Eq. (\ref{asymbc}), then, by Eq. (\ref{ntpwexp}), can
be expressed as%
\begin{equation}
\psi\left(  r,\theta\right)  =\sum_{l=0}^{\infty}\left(  2l+1\right)
i^{l}M_{l}\left(  \frac{i}{kr}\right)  \frac{1}{kr}\sin\left[  kr-\frac{l\pi
}{2}+\Delta_{l}\left(  \frac{i}{kr}\right)  \right]  P_{l}\left(  \cos
\theta\right)  +f\left(  r,\theta\right)  \frac{e^{ikr}}{r}. \label{asymbcexp}%
\end{equation}

\subsection{Scattering wave function}

The scattering wave function can be calculated by imposing the boundary
condition (\ref{asymbcexp}) on the asymptotic wave function (\ref{psiasy}).

Observing the outgoing part of the wave function (\ref{asymbcexp}), $f\left(
r,\theta\right)  e^{ikr}/r$, we can see that the leading contribution of
$f\left(  r,\theta\right)  $ must only be a zero power of $r$, or else the
outgoing wave is not a spherical wave when $r\rightarrow\infty$. Thus, we can
expand $f\left(  r,\theta\right)  $ by the Bessel polynomial, which is
complete and orthogonal \cite{krall1949new}, as
\begin{equation}
f\left(  r,\theta\right)  =\sum_{l=0}^{\infty}g_{l}\left(  \theta\right)
y_{l}\left(  \frac{i}{kr}\right)  . \label{fexpanY}%
\end{equation}
The reason why only $y_{l}\left(  \frac{i}{kr}\right)  $ appears in the
expansion (\ref{fexpanY}) is that only the flux corresponding to $y_{l}\left(
\frac{i}{kr}\right)  e^{ikr}/r$ is an outgoing spherical wave; or, in other
words, the requirement that the scattering wave must be an outgoing wave rules
out the terms including $y_{l}^{\ast}\left(  \frac{i}{kr}\right)
=y_{l}\left(  -\frac{i}{kr}\right)  $.

Equating Eqs. (\ref{psiasy}) and (\ref{asymbcexp}), using the expansion
(\ref{fexpanY}), $\sin\theta=\left(  e^{i\theta}-e^{-i\theta}\right)  /\left(
2i\right)  $, and the orthogonality, and noting that $M_{l}e^{i\Delta_{l}%
}=y_{l}$, we arrive at%
\begin{align}
\frac{1}{2ik}\left[  \left(  2l+1\right)  -A_{l}e^{i\left(  -l\pi/2+\delta
_{l}\right)  }\right]  P_{l}\left(  \cos\theta\right)  +g_{l}\left(
\theta\right)   &  =0,\nonumber\\
\left(  2l+1\right)  e^{il\pi}-A_{l}e^{-i\left(  -l\pi/2+\delta_{l}\right)  }
&  =0. \label{psieqcond}%
\end{align}

Solving these two equations gives%
\begin{align}
A_{l}  &  =\left(  2l+1\right)  e^{il\pi}e^{i\left(  -l\pi/2+\delta
_{l}\right)  },\\
g_{l}\left(  \theta\right)   &  =-\frac{1}{2ik}\left(  2l+1\right)  \left(
1-e^{i2\delta_{l}}\right)  P_{l}\left(  \cos\theta\right)  .
\end{align}

Then, we arrive at
\begin{equation}
f\left(  r,\theta\right)  =\frac{1}{2ik}\sum_{l=0}^{\infty}\left(
2l+1\right)  \left(  e^{2i\delta_{l}}-1\right)  P_{l}\left(  \cos
\theta\right)  y_{l}\left(  \frac{i}{kr}\right)  . \label{frtheta}%
\end{equation}

When taking the limit $r\rightarrow\infty$, the modification factor --- the
Bessel\ polynomial $y_{l}\left(  \frac{i}{kr}\right)  $ --- tends to $1$, and
$f\left(  r,\theta\right)  $ recovers the scattering amplitude in conventional
scattering theory: $f^{\infty}\left(  \theta\right)  =f\left(  \infty
,\theta\right)  =\frac{1}{2ik}\sum_{l=0}^{\infty}\left(  2l+1\right)  \left(
e^{2i\delta_{l}}-1\right)  P_{l}\left(  \cos\theta\right)  $.

The leading is a $p$-wave modification, because the $s$-wave modification is
$y_{0}\left(  x\right)  =1$.

\subsection{Outgoing wave-front surface}

In conventional scattering theory, the observer is at $r\rightarrow\infty$ and
the outgoing wave is a spherical wave. When the observer is at a finite
distance $r$, the outgoing wave-front surface, however, is a surface of
revolution around the incident direction, since for a spherical potential the
outgoing wave must be cylindrically symmetric.

The outgoing wave-front surface is determined by the outgoing flux
$\mathbf{j}^{sc}$ which serves as its surface normal vector. The outgoing flux
is $\mathbf{j}^{sc}=\mathbf{j}-\mathbf{j}^{in}$, where $\mathbf{j=}\frac
{\hbar}{m}\operatorname{Im}\left(  \psi^{\ast}\nabla\psi\right)  $ and
$\mathbf{j}^{in}\mathbf{=}\frac{\hbar}{m}\operatorname{Im}\left(  \psi
^{in\ast}\nabla\psi^{in}\right)  $. Here we write the wave function
(\ref{asymbc}) as $\psi=\psi^{in}+\psi^{sc}$ with $\psi^{in}=e^{ikr\cos\theta
}$ and $\psi^{sc}=f\left(  r,\theta\right)  e^{ikr}/r$.

The outgoing wave-front surface is a surface of revolution. Its generatrix,
$r=r\left(  \theta\right)  $, with $\mathbf{j}^{sc}$ as the normal vector, is
determined by%
\begin{equation}
\frac{1}{r\left(  \theta\right)  }\frac{dr\left(  \theta\right)  }{d\theta
}=-\frac{j_{\theta}^{sc}}{j_{r}^{sc}}=-\tan\gamma^{sc}, \label{rtheta}%
\end{equation}
where $\gamma^{sc}$ is the intersection angle between $\mathbf{j}^{sc}$ and
the radial vector.

The equation of the generatrix, Eq. (\ref{rtheta}), is a differential
equation. The integration constant can be chosen as $r\left(  0\right)  =R$,
where $R$ is the intersection between the outgoing wave-front surface on which
the observer stands and the target along the $z$-axis. Then the solution of
Eq. (\ref{rtheta}) can be formally written as $r=r\left(  \theta,R\right)  $.

Moreover, the Gaussian curvature of the outgoing wave-front surface reads
\begin{equation}
K\left(  \theta\right)  =\frac{1}{r^{2}}\cos^{2}\gamma^{sc}\left(
1+\frac{d\gamma^{sc}}{d\theta}\right)  \left(  1-\tan\gamma^{sc}\cot
\theta\right)  .
\end{equation}
When $r\rightarrow\infty$, $\gamma^{sc}\rightarrow0$, and then $K=1/r^{2}$
reduces to a curvature of a sphere.

\subsection{Differential scattering cross\ section}

The differential scattering section is $d\sigma=\mathbf{j}^{sc}\cdot
d\mathbf{S/}j^{in}$. The scattering flux $\mathbf{j}^{sc}$, other than that in
conventional scattering theory, is not along the radial direction. Thus,%

\begin{equation}
d\sigma=\frac{\mathbf{j}^{sc}\cdot d\mathbf{S}}{j^{in}}=\frac{j^{sc}}{j^{in}%
}\frac{r^{2}d\Omega}{\cos\gamma^{sc}}=\left(  1+\tan^{2}\gamma^{sc}\right)
\frac{j_{r}^{sc}}{j^{in}}r^{2}d\Omega, \label{dsigma}%
\end{equation}
where $j^{sc}=\sqrt{j_{r}^{sc2}+j_{\theta}^{sc2}}$ and $\tan\gamma
^{sc}=j_{\theta}^{sc}/j_{r}^{sc}$. A straightforward calculation gives%
\begin{equation}
\frac{d\sigma}{d\Omega}=\left[  \left\vert f\left(  r,\theta\right)
\right\vert ^{2}+\eta\left(  r,\theta\right)  \right]  \left(  1+\tan
^{2}\gamma^{sc}\right)  , \label{dcs}%
\end{equation}
where%
\begin{equation}
\eta\left(  r,\theta\right)  =\frac{1}{k}\operatorname{Im}\left\{  f^{\ast
}\frac{\partial f}{\partial r}+e^{ikr(1-\cos\theta)}\left\{  \left[
ikr(1+\cos\theta)-1\right]  f+r\frac{\partial f}{\partial r}\right\}
\right\}  .
\end{equation}

\subsection{Total scattering cross section}

For simplicity, we only consider the leading contribution of the total
scattering cross section, $\sigma_{t}\left(  R\right)  =2\pi\int_{0}^{\pi
}\left\vert f\left(  R,\theta\right)  \right\vert ^{2}\sin\theta d\theta$, in
which the outgoing wave-front surface is approximately a sphere of radius $R$.

The total cross section then reads%
\begin{equation}
\sigma_{t}\left(  R\right)  =\frac{4\pi}{k^{2}}\sum_{l=0}^{\infty}\left(
2l+1\right)  \sin^{2}\delta_{l}\left\vert y_{l}\left(  \frac{i}{kR}\right)
\right\vert ^{2}.
\end{equation}
In comparison with conventional scattering theory, a modification factor
$\left\vert y_{l}\left(  \frac{i}{kR}\right)  \right\vert ^{2}$ appears.

\section{Conclusions and outlook \label{conclusions}}

We show that one can obtain a rigorous scattering theory without the
precondition $r\rightarrow\infty$. A rigorous scattering theory contains the
information of the distance between target and observer is presented. The
conventional scattering theory can be recovered by setting $r\rightarrow
\infty$.

In comparison with conventional scattering theory, there is an additional
factor --- the $l$-th Bessel polynomial --- appears in the $l$-th partial-wave
contribution. The leading modification is $p$-wave.

Quantum scattering theory plays an important role in many physical area and is
intensively studied. Nevertheless, all studies are based on conventional
scattering theory. Based on our result, we can\ further consider many
scattering-related problems. For example, at low temperatures, the thermal
wavelength has the same order of magnitude as the interparticle spacing, so
the scattering in a BEC transition \cite{arnold2001bec,kastening2004bose} and
in a transport of spin-polarized fermions
\cite{mullin1983exact,mineev2005theory} may need to take the effect of the
distance into account. The scattering spectrum method is important in quantum
field theory
\cite{rahi2009scattering,forrow2012variable,lambrecht2006casimir,rajeev2011dispersion}%
; a scattering spectrum method without asymptotics can also be discussed.
Moreover, the relation between scattering spectrum method and heat kernel
method, which is given by Ref. \cite{pang2012relation} based on Refs.
\cite{dai2009number,dai2010approach}, can also be improved by the exact result
of the scattering theory without infinite-distance asymptotics. Moreover, a
related inverse scattering problems can also be systematically studied, and
the result can be applied to, e.g., the interference pattern of Bose-Einstein
condensates \cite{liu2000nonlinear} and the Aharonov--Bohm effect
\cite{nicoleau2000inverse}.


\acknowledgments

We are very indebted to Dr G. Zeitrauman for his encouragement. This work is
supported in part by NSF of China under Grant No. 11075115.




\begin{thebibliography}{99}


\bibitem{abramowitz1964handbook}
M.~Abramowitz and I.~A. Stegun, {\em Handbook of Mathematical Functions: With
  Formulars, Graphs, and Mathematical Tables}, vol.~55.
\newblock DoverPublications. com, 1964.

\bibitem{joachain1975quantum}
C.~J. Joachain, {\em Quantum collision theory}.
\newblock North-Holland Publishing Company, Amsterdam, 1975.

\bibitem{krall1949new}
H.~L. Krall and O.~Frink, {\it A new class of orthogonal polynomials: The
  bessel polynomials},  {\em Transactions of the American Mathematical Society}
  {\bf 65} (1949), no.~1 100--115.

\bibitem{arnold2001bec}
P.~Arnold and G.~Moore, {\it Bec transition temperature of a dilute homogeneous
  imperfect bose gas},  {\em Physical Review Letters} {\bf 87} (2001), no.~12
  120401.

\bibitem{kastening2004bose}
B.~Kastening, {\it Bose-einstein condensation temperature of a homogenous
  weakly interacting bose gas in variational perturbation theory through seven
  loops},  {\em Physical Review A} {\bf 69} (2004), no.~4 043613.

\bibitem{mullin1983exact}
W.~Mullin and K.~Miyake, {\it Exact transport properties of degenerate, weakly
  interacting, and spin-polarized fermions},  {\em Journal of low temperature
  physics} {\bf 53} (1983), no.~3-4 313--338.

\bibitem{mineev2005theory}
V.~Mineev, {\it Theory of transverse spin dynamics in a polarized fermi liquid
  and an itinerant ferromagnet},  {\em Physical Review B} {\bf 72} (2005),
  no.~14 144418.

\bibitem{rahi2009scattering}
S.~J. Rahi, T.~Emig, N.~Graham, R.~L. Jaffe, and M.~Kardar, {\it Scattering
  theory approach to electrodynamic casimir forces},  {\em Physical Review D}
  {\bf 80} (2009), no.~8 085021.

\bibitem{forrow2012variable}
A.~Forrow and N.~Graham, {\it Variable-phase s-matrix calculations for
  asymmetric potentials and dielectrics},  {\em Physical Review A} {\bf 86}
  (2012), no.~6 062715.

\bibitem{lambrecht2006casimir}
A.~Lambrecht, P.~A.~M. Neto, and S.~Reynaud, {\it The casimir effect within
  scattering theory},  {\em New Journal of Physics} {\bf 8} (2006), no.~10 243.

\bibitem{rajeev2011dispersion}
S.~Rajeev, {\it A dispersion relation for the density of states with
  application to the casimir effect},  {\em Annals of Physics} {\bf 326}
  (2011), no.~6 1536--1547.

\bibitem{pang2012relation}
H.~Pang, W.-S. Dai, and M.~Xie, {\it Relation between heat kernel method and
  scattering spectral method},  {\em The European Physical Journal C} {\bf 72}
  (2012), no.~5 1--13.

\bibitem{dai2009number}
W.-S. Dai and M.~Xie, {\it The number of eigenstates: counting function and
  heat kernel},  {\em Journal of High Energy Physics} {\bf 2009} (2009), no.~02
  033.

\bibitem{dai2010approach}
W.-S. Dai and M.~Xie, {\it An approach for the calculation of one-loop
  effective actions, vacuum energies, and spectral counting functions},  {\em
  Journal of High Energy Physics} {\bf 2010} (2010), no.~6 1--29.

\bibitem{liu2000nonlinear}
W.-M. Liu, B.~Wu, and Q.~Niu, {\it Nonlinear effects in interference of
  bose-einstein condensates},  {\em Physical Review Letters} {\bf 84} (2000),
  no.~11 2294.

\bibitem{nicoleau2000inverse}
F.~Nicoleau, {\it An inverse scattering problem with the aharonov--bohm
  effect},  {\em Journal of Mathematical Physics} {\bf 41} (2000) 5223.










\end{thebibliography}
\end{document}